\RequirePackage{fix-cm}
\documentclass{svjour3}                     
\smartqed  
\usepackage{graphicx}
\usepackage{multirow}
\usepackage{subcaption}
\usepackage{tabularx}
\usepackage{xparse}
\usepackage{mathtools}
\newsavebox{\mybox}
\NewDocumentEnvironment{strange}{}
    {%
        \lrbox\mybox
        \tabularx{0.8\textwidth}{|l|X|}%
    }
    {%
        \endtabularx
        \endlrbox
    }

\DeclarePairedDelimiter\abs{\lvert}{\rvert}

\usepackage{upquote}
\def\hat{\mathaccent "705E\relax}

\usepackage{times}
\journalname{Int J CARS}

\begin{document}

\title{The Effect of Video Playback Speed on Surgeon Technical Skill Perception}
 


\author{Jason D. Kelly         \and
	     Ashley Petersen Ph.D. \and
	     Thomas S. Lendvay M.D. \and
          Timothy M. Kowalewski Ph.D.
}


\institute{Jason Kelly \at
              Department of Mechanical Engineering, University of Minnesota, Minneapolis, MN \\
              \email{kell1917@umn.edu}           
		\and
		A. Petersen \at
		Division of Biostatistics, University of Minnesota, Minneapolis, MN
		\and 
		T. Lendvay \at
		Department of Urology, Seattle Children's Hospital, Seattle, WA
           \and
           T. Kowalewski \at
              Department of Mechanical Engineering, University of Minnesota, Minneapolis, MN
}

\date{Received: date / Accepted: date}

\maketitle

\begin{abstract}
\textit{Purpose:} Finding effective methods of discriminating surgeon technical skill has proven a complex problem to solve computationally. Previous research has shown that obtaining non-expert crowd evaluations of surgical performances is as accurate as the gold standard, expert surgeon review \cite{CSATS_Paper}. The aim of this research is to learn whether crowd-sourced evaluators give higher ratings of technical skill to video of performances with increased playback speed, its effect in discriminating skill levels, and whether this increase is related to the evaluator consciously being aware that the video is being manually manipulated. \\ 
\textit{Methods:} A set of ten peg transfer videos (5 novices, 5 experts), were used to evaluate the perceived technical skill of the performers at each  video playback speed used (0.4x-3.6x). Objective metrics used for measuring technical skill were also computed for comparison by manipulating the corresponding kinematic data of each performance. Two videos of an expert and novice performing dry lab laparoscopic trials of peg transfer tasks were used to obtain evaluations at each playback speed (0.2x-3.0x) of perception of whether a video is played at real-time playback speed or not. \\
\textit{Results:}  We found that while both novices and experts had increased perceived technical skill as the video playback was increased, the amount of increase was significantly greater for experts. Each increase in the playback speed by 0.4x was associated with, on average, a 0.72-point increase in the GOALS score (95\% CI: 0.60-0.84 point increase; p $<$ 0.001) for expert videos and only a 0.24-point increase in the GOALS score (95\% CI: 0.13-0.36 point increase; p $<$ 0.001) for novice videos.\\
\textit{Conclusion:} Due to the differential increase in perceived technical skill due to increased playback speed for experts, the difference between novice and expert skill levels of surgical performances may be more easily discerned by manually increasing the video playback speed. \\
\end{abstract}

\section{Introduction}
\label{intro}

Medical errors make up a third of all deaths in the United States, one of the largest contributors of which are surgical errors \cite{ThirdLeadingDeath}. Technical surgical skill is directly related to patient outcomes \cite{Birkmeyer}, but it remains a difficult computational task to correctly classify surgeons into skill levels with a compelling level of accuracy, i.e. never misclassifying an `obvious novice' as an `obvious expert' and vice versa - the MAC Criterion \cite{MAC}. The de facto gold standard for evaluating technical skill is video evaluation by an expert surgeon using Likert-scale assessment metrics, in which evaluators submit ratings on an anchored scale of 1-5. Using crowds of \textit{non-expert} evaluators are a surprisingly accurate way to inexpensively and rapidly obtain skill level ratings for videos of surgical performances, with a pass/fail rate capable of matching 100\% of ratings by expert surgeons \cite{CSATS}. The fact remains, however, that humans can be biased in their thinking, and subjective metrics of rating performances can lead to results we would not expect from computational models of evaluation. 

\begin{figure}
\centering
	\includegraphics[width = 0.6\textwidth]{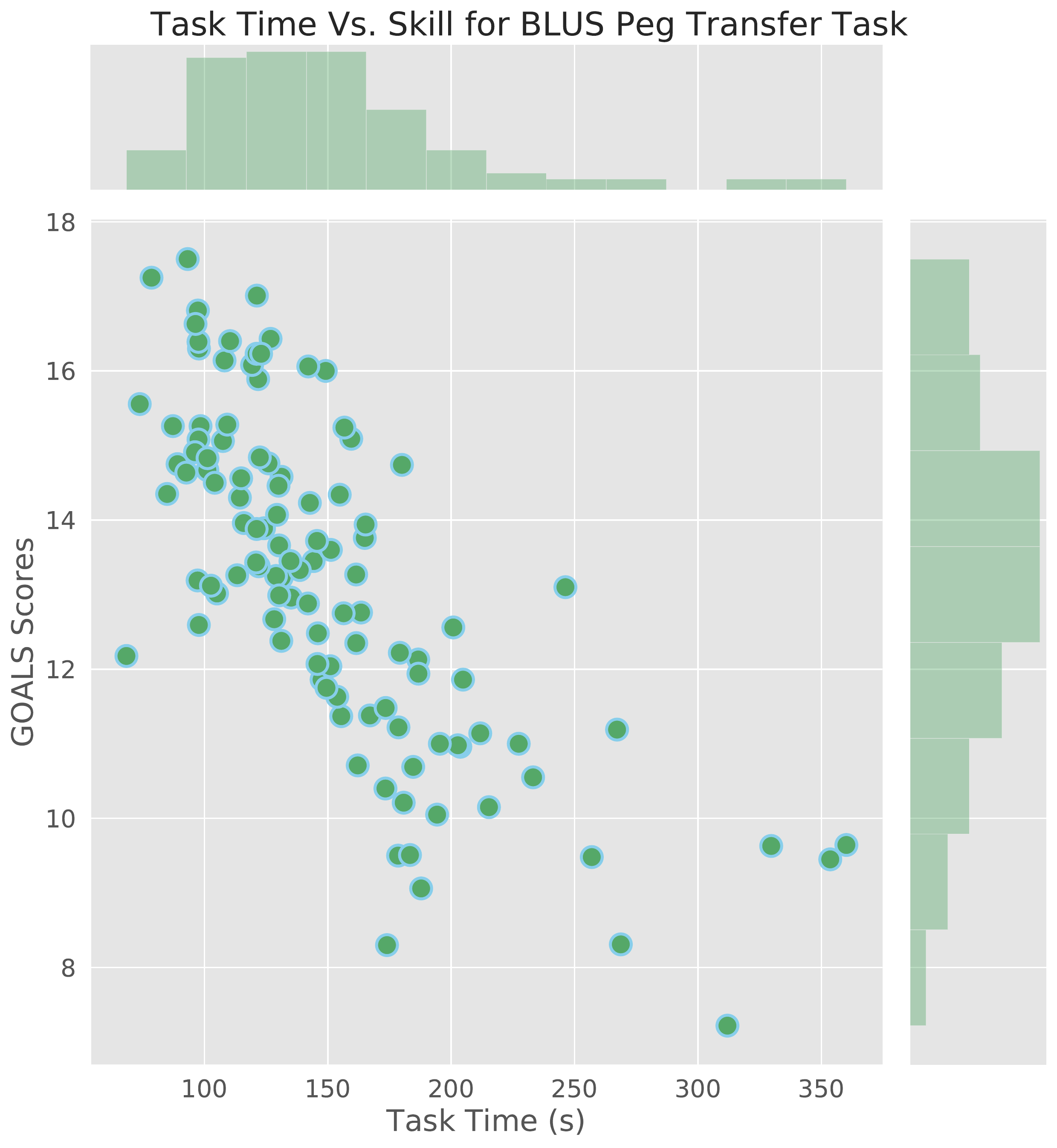}
	\caption{Task time completion vs. GOALS scores for all peg transfer performances in the Basic Laparoscopic Urologic Skills (BLUS) dataset \cite{Initial_BLUS_curriculum}.}
	\label{fig:Task_Time}
\end{figure}

One of the most popular laparoscopic surgical skill assessment metrics is the Global Operative Assessment of Laparoscopic Skills (GOALS), which is the most common objective assessment tool for laparoscopic performance studies \cite{GOALS_Paper1}, \cite{GOALS_Paper2}. The subdomains in this metric include: bimanual dexterity, tissue handling, efficiency, depth perception, and autonomy. Task time is not a direct metric used to evaluate the technical skill of laparoscopic surgeons with tools like GOALS, however time for task completion is often seen as one of the most predictive objective forms of evaluating technical skill, as seen in Fig. \ref{fig:Task_Time} \cite{BeyondTaskTime}, \cite{Hung}.

\begin{figure}
	\centering
	\includegraphics[scale=0.35]{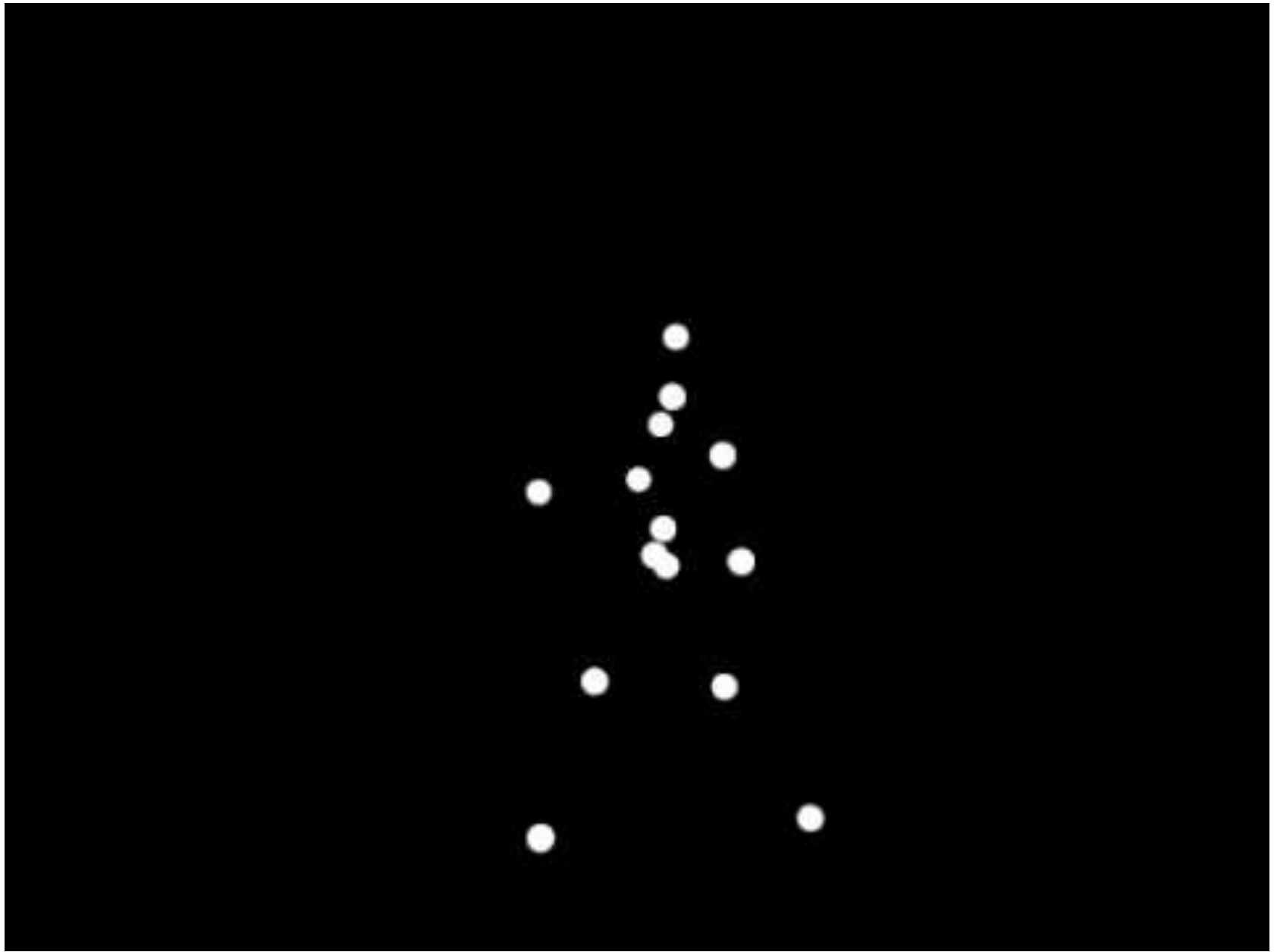}
	\caption{A point light walker, used in biological motion research. Despite highly impoverished data (points only, no video), humans seem to quickly infer human gait within a few steps of animation and even extract subtle cues like gender or emotional state \cite{PLW}, \cite{altered_PLW}.}
	\label{fig:PLW}
\end{figure}

\subsection{Biological Motion Perception}
\label{BiologicalMotionPerception}

Previous research involving point light walkers (PLW), 10-12 dots of light illuminated on a screen to correspond to the human joints in walking motion, (illustrated in Fig. \ref{fig:PLW}), revealed that humans are excellent at recognizing more subtle qualitative characteristics like the gender or emotion of a PLW by inferring the gait from a few moving points \cite{PLW}. Once the PLW's walking gait speed is changed however, the ability to recognize gender is diminished \cite{altered_PLW}. This and other work suggest that biological motion perception is somehow tuned to the speed of the motion involved \cite{speed_bioMotion}. 

Due to task time being such a strong indicator of surgical technical skill, we speculate about whether a video artificially sped up to simulate a quicker time for task completion will lead to an altered perception of technical skill, or if human raters evaluate other, more nuanced, qualities of the performance.

\subsection{Objective Metrics}
\label{Objective Metrics}

Other metrics exist for measuring technical skill, with varying degrees of success at classifying novice and expert skill levels. These techniques utilize kinematic tool tip data, as opposed to using video frames to make measurements. Metrics which use algorithms or computational methods to evaluate performance are sometimes referred to as Automated Performance Metrics (APMs) \cite{Hung}. The simplest metric to measure, and often the most accurate, is time of task completion. The following APMs seek to compute information more complex than speed to account for technical skill, by conversely using the form of movements made.

Prior work has discovered that recovering stroke patients have a decrease in `jerky' non-smooth movements as they progress in rehabilitation therapy \cite{Jerk_Hogan}. The jerk cost of all tool movements performed in a task has also been used to measure accuracy and skill in areas in which skill is needed \cite{Locations_BLUS}. The jerk cost is computed by taking the integral of the squared jerk summed for both the left and right hand over the course of the performance, shown in Equation \ref{eqn:jerk}, in which $T$ is total time of task performance and $x$ is the magnitude of the movement. 

\begin{center}
\begin{equation}
\textrm{Jerk Cost}= \int_{0}^{T}  \abs{\dddot{x}(t)\hspace{0pt}}^{2}dt 
\label{eqn:jerk}
\end{equation}
\end{center}

Spectral Arc length (SAL) is also related to smoothness of a movement, and aims to use the tangential velocity of hand movements to compute the overall smoothness of a task. This has been used to differentiate skill levels of medical professionals in the past \cite{Omalley_SAL}, \cite{Balas_SAL_Creator}.  To compute the SAL, the motion in a task must be segmented into specific hand grasping movements, with the corresponding speeds and durations known for each grasp. The Fourier magnitude spectrum transform of the speed profile is then computed and normalized with respect to its zero frequency value. The smoothness is then measured for each segmented grasp, shown in Equation \ref{eqn:SAL}, in which $\omega$ is frequency, $\omega_{c}$ is cutoff frequency, and $V$ is the Fourier magnitude spectrum of speed. Finally a weighted average is taken across all grasping motions to arrive at an SAL metric for the task as a whole. 

\begin{equation}
SAL = 
	-\int_{0}^{\omega_{c}}  
	\left[
		\left( \frac{1}{\omega_{c}} \right)^{2} +
		 \left( \frac{d \hat{V}(\omega)}{d \omega} \right)^{2}
	\right]^{\frac{1}{2}} d\omega,$$  $$
	 \hat{V}(\omega) = \frac{V(\omega)}{V(0)}
\label{eqn:SAL}
\end{equation}

Instead of relying on the form of a grasping movement, taking the sum of all movements which occur is an additional metric that can be used to evaluate skill in some domains in which it is believed that novice performers on average perform more grasping movements than experts. A movement can be calculated in several ways, but the most common way of recording is to define a threshold of speed at which when the performer's speed falls below that threshold, a movement has ended and a new grasping motion starts once the magnitude of the velocity has risen above that threshold again. Counting each time this threshold is passed gives this total count of all movements conducted during the task \cite{Kowalewski_Thesis}. 

All of these APMs which may be used to compute the skill of a surgeon have in common that once the speed playback of the data is manipulated, the results normally do not lead to any more separation between skill levels, as changing the magnitude of speed will not affect calculations to a large degree. 

Our motivation in this work is to learn whether, by increasing video playback speed, novices and experts will be perceived by crowds as more skilled when they appear to be moving faster, \textit{and} if experts and novices will have different rates of perceived change as the playback speed is increased. We will do this by evaluating the ability to discriminate obvious novices from experts of both human raters and popular tool motion metrics (APMs), at different playback speeds. We expect each APM's ability to discriminate skill to operate as the experimental control in this study, with a negligible change in separation between skill levels as speed is increased. Finally, we seek to investigate how an evaluator's likely conscious awareness of whether a video is sped up or slowed down relates to their perception of skill. 

\begin{table}[h!]
\caption{Likert-scale technical skill perception questionnaire for a manually sped-up video, from four domains of the GOALS assessment metric.}
\label{tab:GOALS_Questions}
\begin{strange}\hline
Score & \textbf{Depth Perception}  \\ \hline \hline
(1) & Constantly overshoots target, wide swings, slow to correct    \\ \hline
(2) &   \\ \hline
(3) & Some overshooting or missing of target, but quick to correct \\ \hline
(4) &  \\ \hline
(5) & Accurately directs instruments in the correct plane to target \\ \hline
\end{strange}
\begin{center}
\usebox\mybox
\end{center}
\begin{strange}\hline
Score & \textbf{Bimanual Dexterity}\\ \hline \hline
(1) & Uses only one hand, ignores non-dominant hand, poor\\
    & coordination between hands\\ \hline
(2) &   \\ \hline
(3) & Uses both hands, but does not optimize interaction \\
    & between hands \\ \hline
(4) & \\ \hline
(5) & Expertly uses both hands in a complementary manner to\\
    & provide optimal exposure \\ \hline
\end{strange}
\begin{center}
\usebox\mybox
\end{center}
\begin{strange} \hline
Score & \textbf{Efficiency} \\ \hline \hline
(1) & Uncertain, inefficient efforts; many tentative movements; \\ 
    &   constantly changing focus or persisting without progress \\ \hline
(2) &   \\ \hline
(3) & Slow, but planned movements are reasonably organized  \\ \hline
(4) &   \\ \hline
(5) & Confident, efficient and safe conduct, maintains focus \\
    & on task until it is better performed by way of an \\
    & alternative approach \\ \hline
\end{strange}
\begin{center}
\usebox\mybox
\end{center}
\begin{strange} \hline
Score & \textbf{Tissue Handling}\\ \hline \hline
(1) & Definitely sped-up video \\ \hline
(2) &   \\ \hline
(3) & Handles tissue reasonably well, minor trauma to adjacent\\
    & tissue(i.e. occasional unnecessary bleeding or slipping of\\
    & the grasper) \\ \hline
(4) &   \\ \hline
(5) & Handles tissues well, applies appropriate traction, \\
    & negligible injury to adjacent structures \\ \hline
\end{strange}
\begin{center}
\usebox\mybox
\end{center}
\end{table}

\begin{table}[h!]
\centering
\caption{Likert-scale speed perception questionnaire for a manually sped-up video.}
\label{tab:speed_perception_fast}       
\begin{tabular}{ll}
\hline\noalign{\smallskip}
Score & \textbf{Does this video appear to}\\
 & \textbf{ have been altered to have an increased playback speed?} \\ \hline
(1) & Definitely sped-up video \\ 
(2) & Likely sped up video  \\ 
(3) & Not sure / I don't know \\ 
(4) & Not likely sped up video\\ 
(5) & Definitely real-time playback speed (unaltered speed) \\ 
\noalign{\smallskip}\hline
\end{tabular}
\end{table}

\section{Methods}
\label{Methods}
\subsection{Dataset}
\label{Dataset}

This study used the Basic Laparascopic Urologic Study (BLUS) dataset, described in detail in \cite{CSATS_Paper}. This dataset arose from a gap in the field, in which no educational surgical certification process existed for urologic surgery, as opposed to how the Fundamentals of Laparosocopic Surgery (FLS) exists for general surgical procedures \cite{FLS1}, \cite{FLS2}, \cite{FLS3}. The BLUS training curriculum aimed to address urology appropriate skills improvement by recording video performances in an initial validation project of over 450 videos \cite{Initial_BLUS_curriculum}. 

This dataset contains 454 videos of surgical performances consisting of four surgical tasks (110 peg transfer, 110 pattern cutting, 115 suturing, 119 clip applying), which are performed by medical students, urology residents, fellows and faculty surgeons from eight academic urology training centers in the United States \cite{Locations_BLUS}. Each trial of a surgeon performing one of the four tasks was recorded at 30 fps with a fixed camera-position of the laparoscopic tools interacting with the training field. Each trial additionally has kinematic data, sampled at 30 Hz, logging the tooltip positions, grasping force, and the jaw angles during the performance, as well as demographic information for each performer being obtained. A GOALS score for each performance was also previously obtained from either expert or crowd evaluation. 

In previous research the Peg Transfer task has been shown to be one of the most easily differentiable tasks for surgical technical skill. Although the Peg Transfer task is criticized as the least clinically relevant, its clear ability to separate novice/expert skill levels was preferred to explore the research questions herein. Ten videos from the peg transfer task were used, in which five `obvious experts' and five `obvious novices' were chosen as baseline definitions of skill \cite{MAC}. Here an obvious novice is defined as someone who should never be allowed to operate and obvious experts as surgeons who should never be disqualified from operating. An obvious expert was chosen such that the performer was in the top 20\% of experience levels (attending surgeon or faculty urologist), previously obtained GOALS scores, and task completion times of all peg transfer tasks. The obvious novices were chosen in the same fashion such that they were in the bottom quintile of these domains. This method aimed to provide two well discriminated clusters of skill levels. 

Amazon Mechanical Turk was the crowd-sourcing platform used for this study, in which each non-expert crowd worker was paid an average of \$0.10 to watch and evaluate a video, depending on the video duration. A web domain was created for which Turkers would be redirected to, where they submitted a consent form and were asked questions about videos. Two different kinds of experiments were conducted: \textit{technical skill perception} and \textit{speed perception}.

\subsection{Experiment 1: Technical Skill Perception}
\label{Experiment1}

Technical skill perception was measured by surveying non-expert crowds to give each video performance a GOALS score by rating each of the 4 subdomains shown in Table \ref{tab:GOALS_Questions}. Forty ``turkers" were recruited per video, in which each video at each playback speed was independently submitted to the website in order to avoid a grouping bias. Videos were altered to speeds in the range 0.4x-3.6x, moving in intervals of 0.4, edited using FFmpeg \cite{FFMPEG}, in which frames were either taken out or added in order to create the resulting playback speed. The score from each of four subdomains (Depth Perception, Bimanual Dexterity, Efficiency, and Tissue Handling) were summed to create a cumulative score for each performance in the range of 4-20. The mean of each video's cumulative GOALS score was recorded, with a 95\% confidence interval.

\subsection{Experiment 2: Speed Perception}
\label{Experiment2}

The video playback speed perception test was conducted by using one novice and one expert from the peg transfer task, and asking the evaluator if they thought the video was being manually edited. If we showed them a video with a decreased playback speed, the worker was asked whether they thought the video was running at real-time speed or was manually edited to be slowed down. In the same way, workers were asked if the video was being sped up or played at real-time playback speed for videos which had an increased video playback speed. These Likert-scale questions are shown in Table \ref{tab:speed_perception_fast} for a video in which the playback speed was increased. Speeds tested ranged from 0.2x-3.0x, moving in intervals of 0.2x to verify the range of speeds to use for the technical skill perception tests.

\subsection{APM Validation at Each Playback Speed}
The APMs mentioned in Section \ref{Objective Metrics} were also recorded at these various playback speeds for comparison to crowd evaluations. These were recalculated by scaling the magnitude of the speeds in the kinematic data to be equivalent to the speed displayed in a given video playback speed, then computing these objective metrics. For the jerk cost, this involved using a Holoborodko smooth noise-robust differentiator \cite{Holoborodko}. For the others this involved simply scaling the magnitude of speed and inputting the result to the pre-created function for each corresponding APM. All APMs were computed in Python and MATLAB, \cite{Python}, \cite{MATLAB}.


\section{Results}

\subsection{Technical Skill Perception}

\begin{figure}[h!]
  \centering
  \hspace{-0.8cm}
  \includegraphics[width=0.99\textwidth]{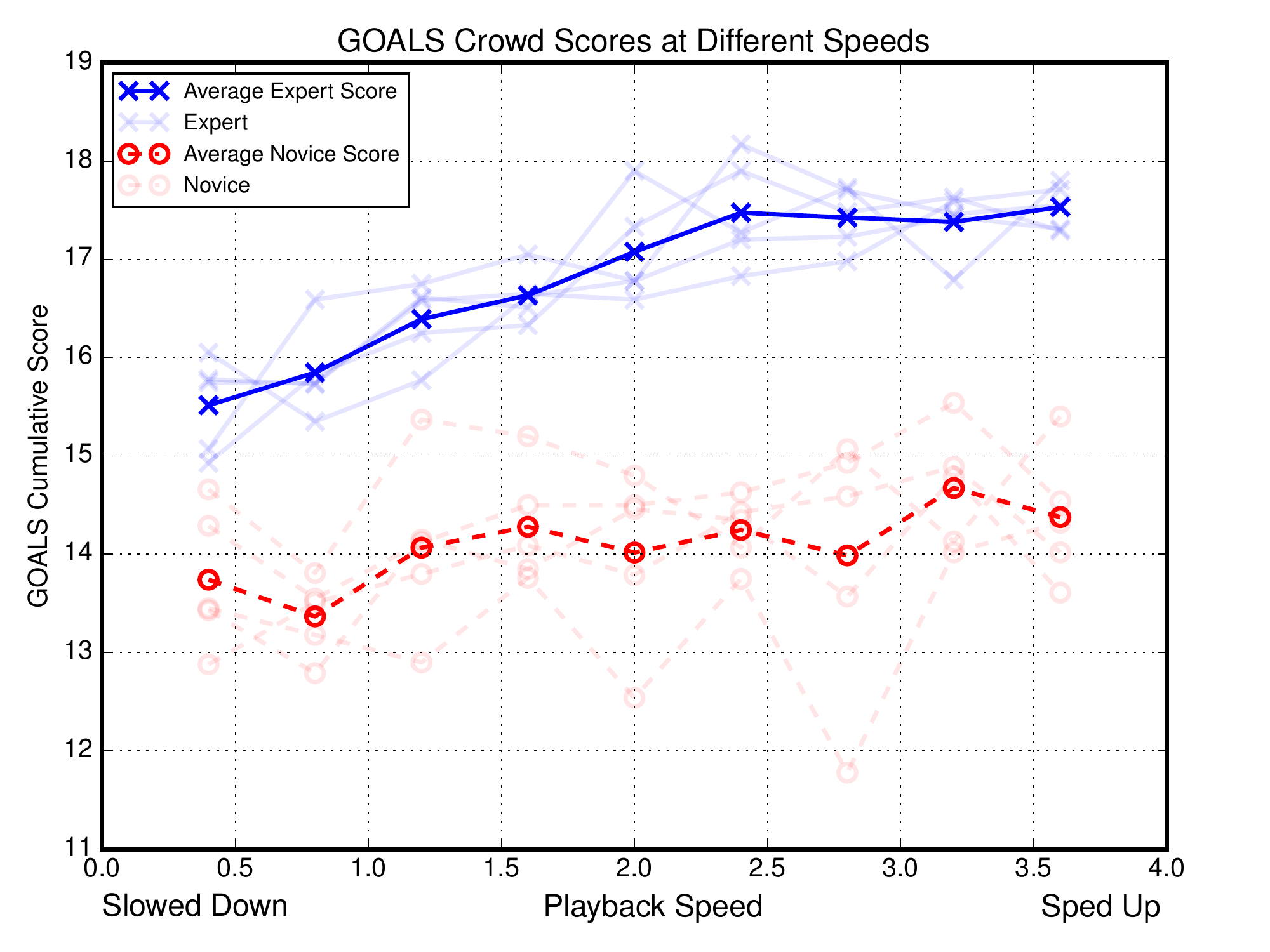}
\caption{All mean crowd evaluations from each novice and expert peg transfer task at various video playback speeds. (Each solid marker indicates N = 40).}
\label{fig:All_Scores}       
\end{figure}

\begin{figure}[h!]

	\centering
	
    \begin{subfigure}[t]{0.49\textwidth}
    
        \raisebox{-\height}{\includegraphics[width=\textwidth]{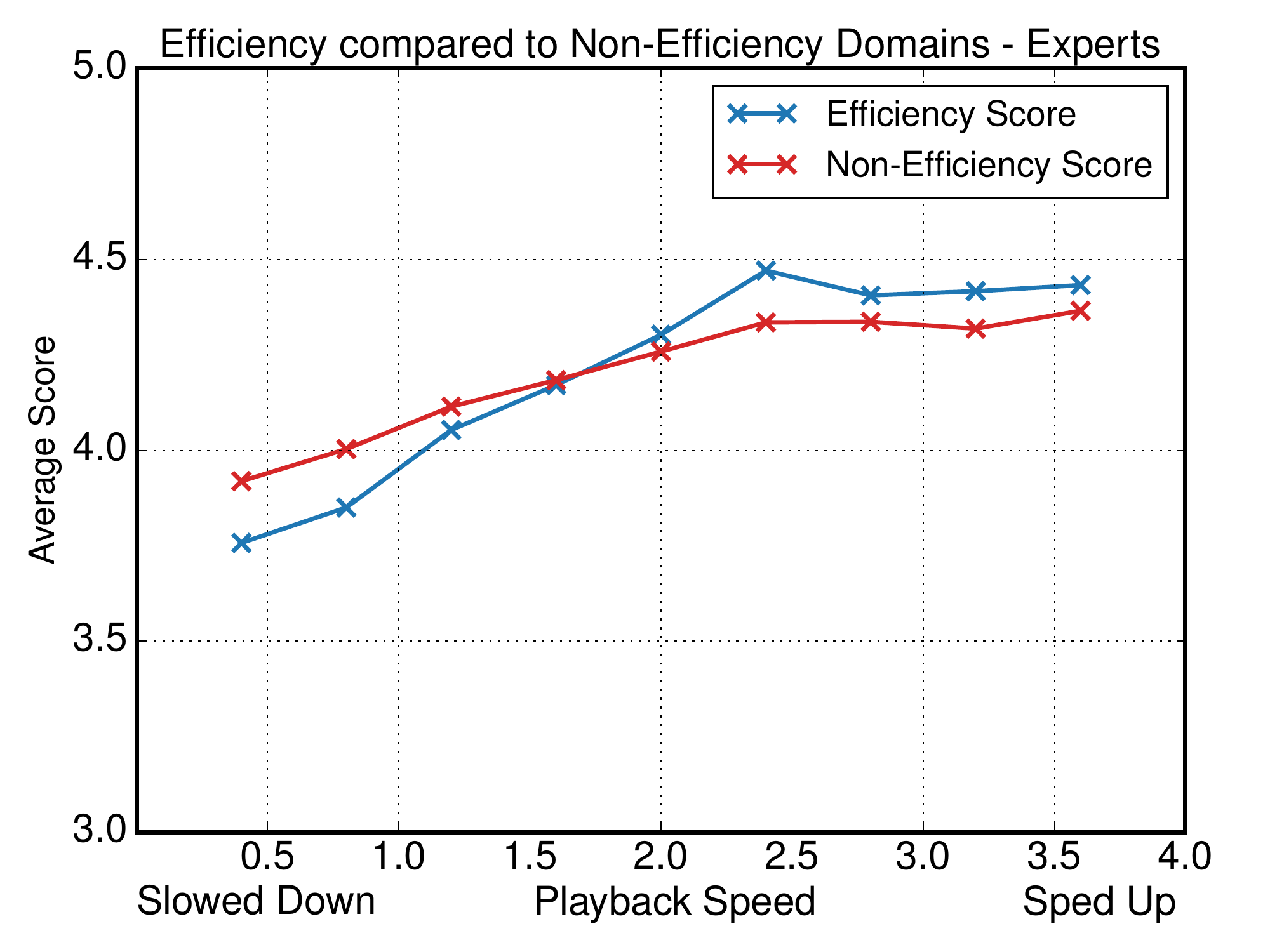}}
        
    \end{subfigure}
    \begin{subfigure}[t]{0.49\textwidth}
        \raisebox{-\height}{\includegraphics[width=\textwidth]{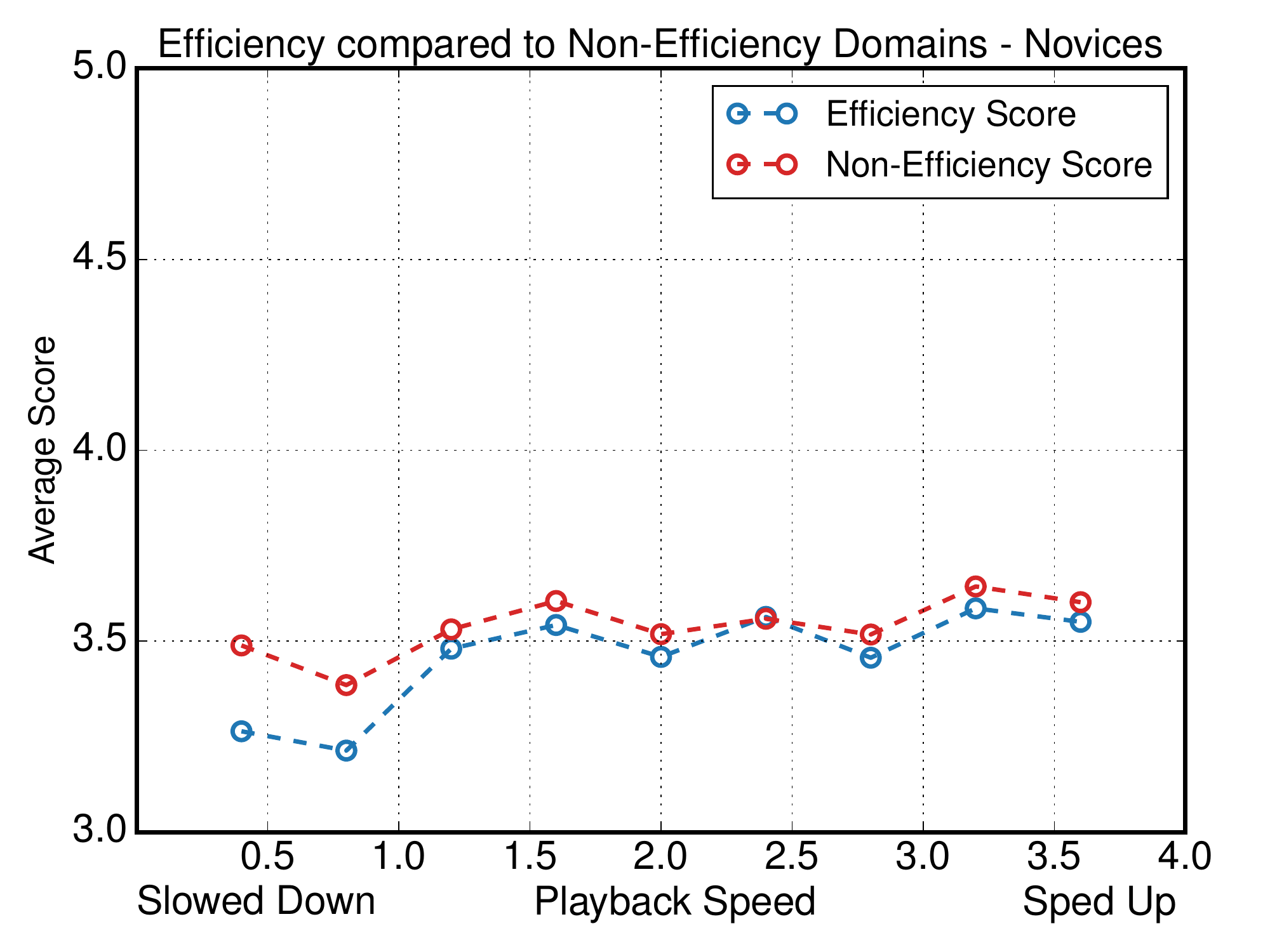}}
    \end{subfigure}
	\caption{The efficiency subdomain as compared to the average of the other GOALS subdomains, for experts and novices.}
	\label{fig:Stacked_Area}
	
\end{figure}
The mean of the GOALS score for each obvious expert and novice are shown in Fig. \ref{fig:All_Scores}. For expert videos, each increase in the playback speed by 0.4x was associated with, on average, a 0.72-point increase in the GOALS score (95\% CI: 0.60-0.84 point increase; p $<$ 0.001). On average these scores appear to increase within a sublevel of the playback speeds around 0.5x to 2.4x, and then level out at all remaining playback speeds. For novice videos, each increase in the playback speed by 0.4x was associated with, on average, a 0.24-point increase in the GOALS score (95\% CI: 0.13-0.36 point increase; p $<$ 0.001). Thus, while both experts and novices had increased perceived technical skill as the playback speed was increased, the gain was significantly greater for experts. The experts had, on average, a 0.47-point greater increase (95\% CI: 0.31-0.64; p $<$ 0.001) in the GOALS score, compared to novices, for each 0.4x increase in playback speed. Fig. \ref{fig:Stacked_Area} shows the increase in the efficiency subdomain as well as the average of the other three domains to visualize whether efficiency (seen as the most related to speed) is the only increasing domain. As shown, it is clear that the other domains increased at nearly the same rate. 

\begin{figure}[h!]
     \centering
    \begin{subfigure}[t]{0.49\textwidth}
        \raisebox{-\height}{\includegraphics[width=1.02\textwidth]{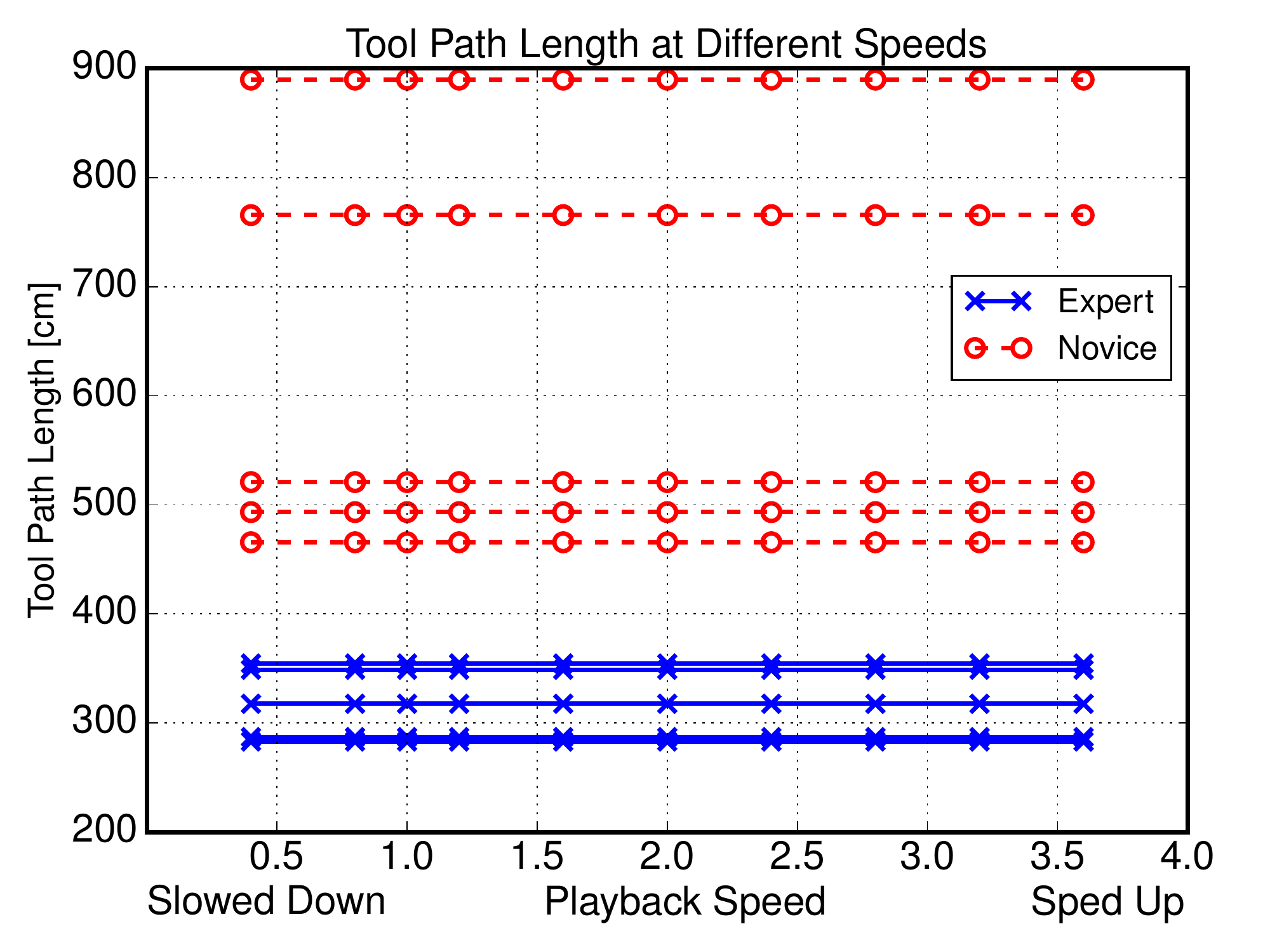}}
        
    \end{subfigure}
    \hfill
    \begin{subfigure}[t]{0.49\textwidth}
        \raisebox{-\height}{\includegraphics[width=1.02\textwidth]{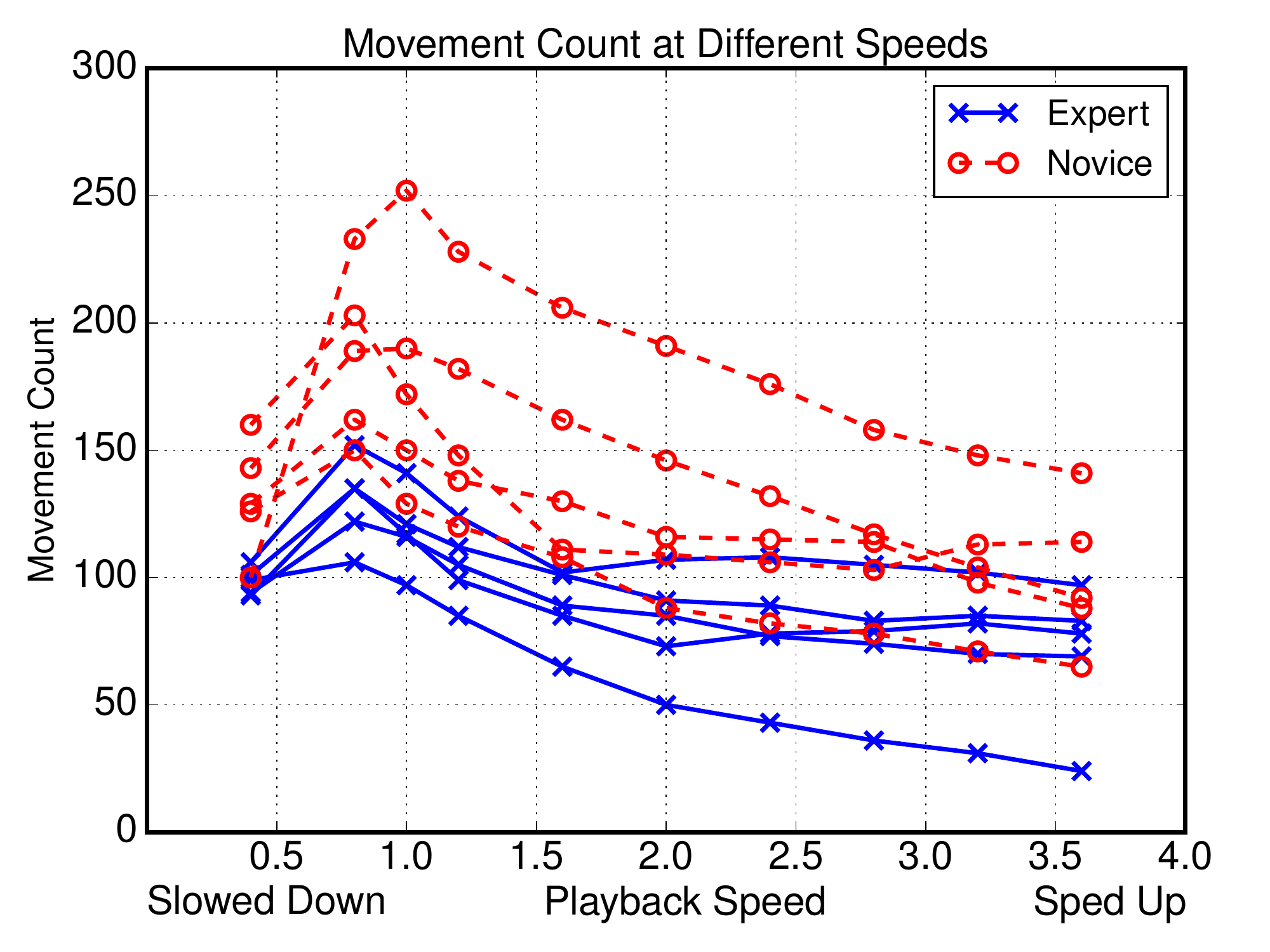}}
       
    \end{subfigure}
    \begin{subfigure}[t]{0.49\textwidth}
        \raisebox{-\height}{\includegraphics[width=1.02\textwidth]{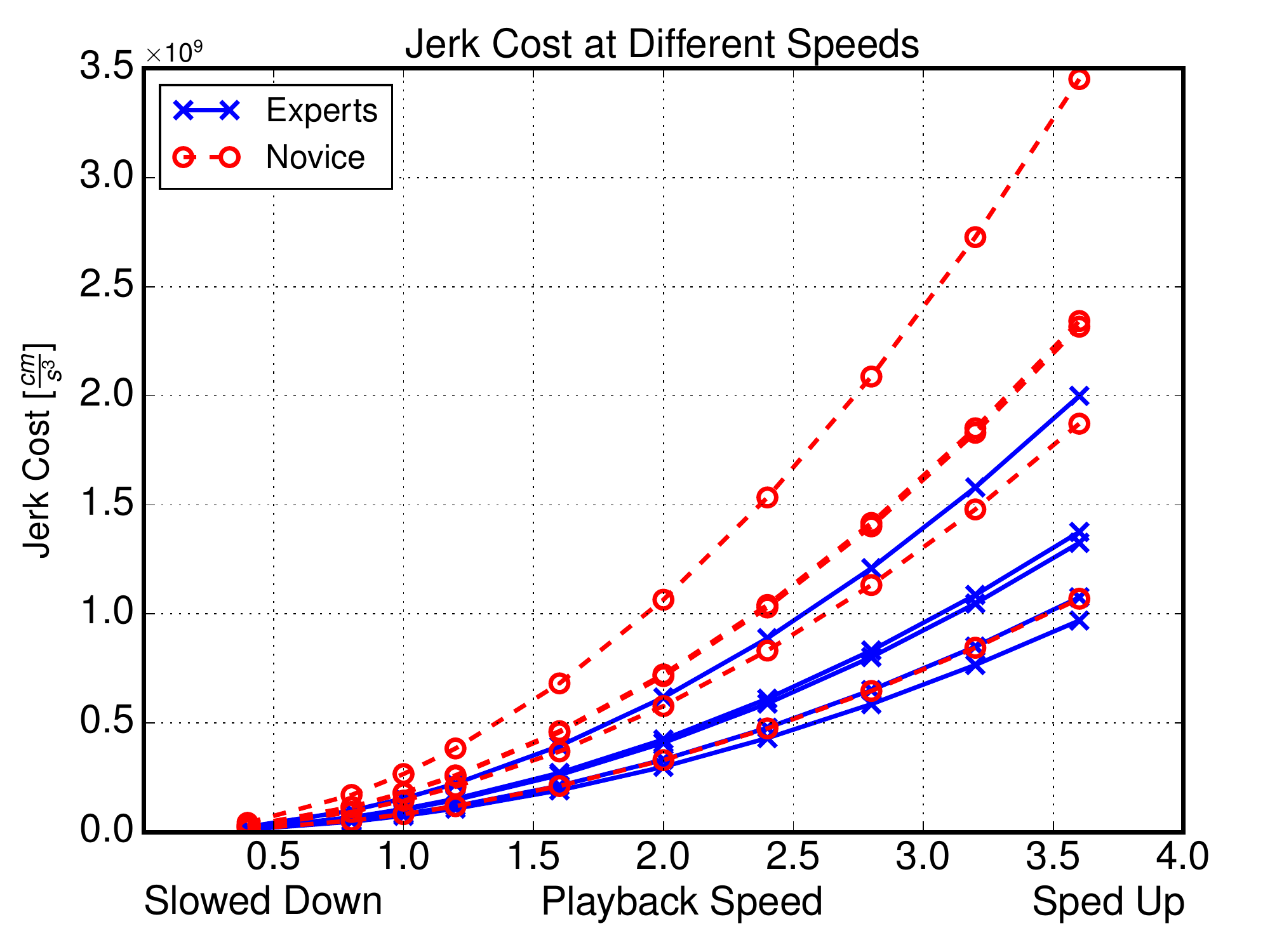}}
        
    \end{subfigure}
    \hfill
    \begin{subfigure}[t]{0.49\textwidth}
        \raisebox{-\height}{\includegraphics[width=1.02\textwidth]{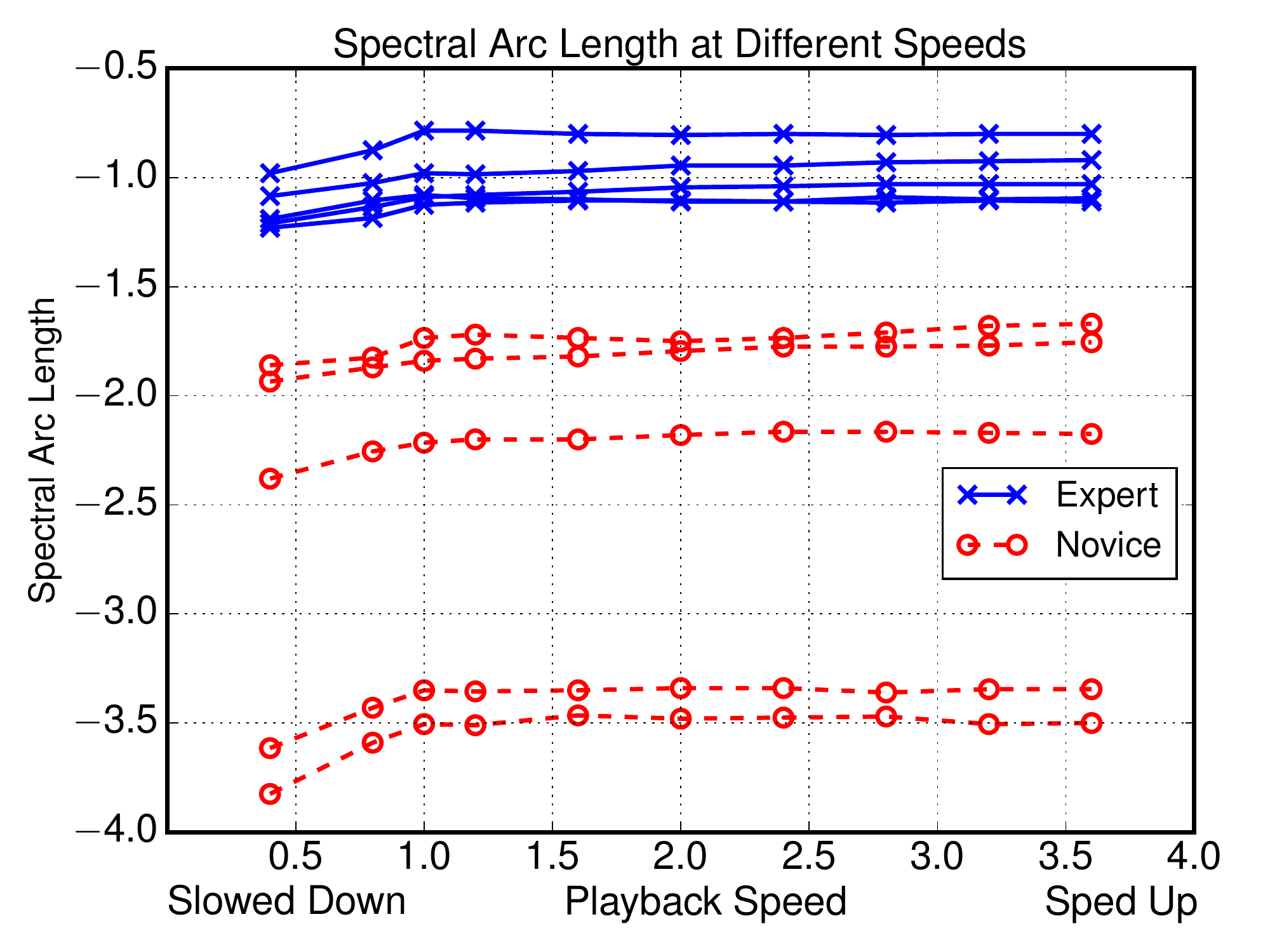}}
       
    \end{subfigure}
    \caption{Objective technical skill metrics for the obvious experts and novices across various speeds.}
    \label{fig:Objective_Metrics_Plot}
\end{figure}

\begin{figure}[h!]

	\includegraphics[width=\textwidth]{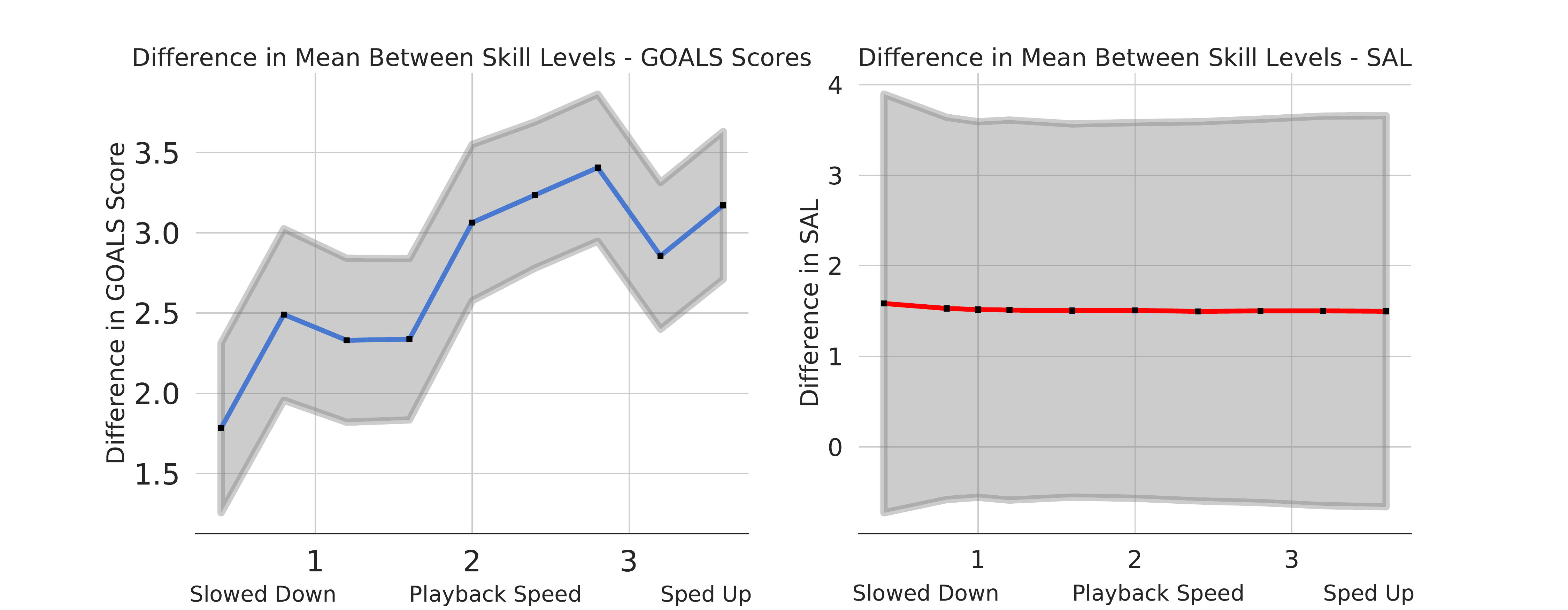}
    \caption{Difference in mean between skill levels from the highest performing APM and from crowd scores, with a 95\% CI in the shaded regions.}
    \label{fig:Difference_in_Mean_Objective}
\end{figure}

The APMs which were calculated with the manipulated kinematic data are shown in Fig. \ref{fig:Objective_Metrics_Plot}. As shown, most metrics show very little change at each of the different speeds, but even if they change, performers' technical skills are not able to be discriminated between expert and novice skill any more easily. These metrics were also calculated by incorporating a combination of scaling the speeds and removing indices corresponding to the frames which would be removed from videos when the playback speed is increased. These figures aren't included, but the results exhibit the same lack of separation between the skill groups at various playback speeds. Fig. \ref{fig:Difference_in_Mean_Objective} shows the difference in mean for experts and novices, tested with both the crowd scores and the highest performing APM, spectral arc length, to illustrate the difference in skill discrimination between crowd workers and the most accurate APM as speed is increased. 

\begin{figure}[h!]
  \begin{subfigure}[t]{\textwidth}
  \centering
  	\raisebox{-\height}{\includegraphics[width=0.6\textwidth]{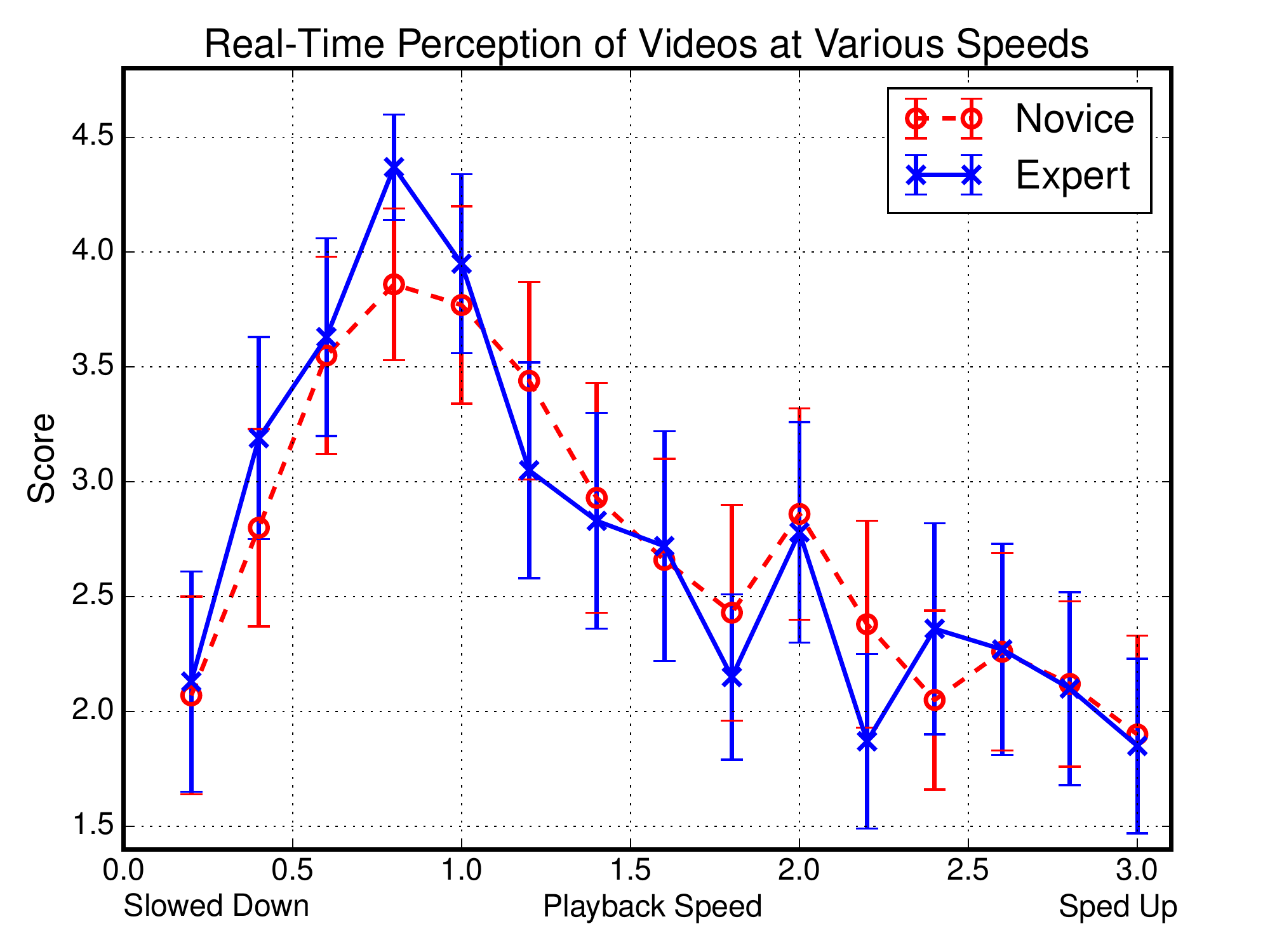}}
	\caption{All mean crowd evaluations of video speed perception at various video playback speeds. (Each solid marker indicates N = 40).}
	\label{fig:Speed_Perception_Scores}
	\end{subfigure}
	\hfill
	\begin{subfigure}[t]{\textwidth}
	\centering
        \raisebox{-\height}{\includegraphics[width=0.6\textwidth]{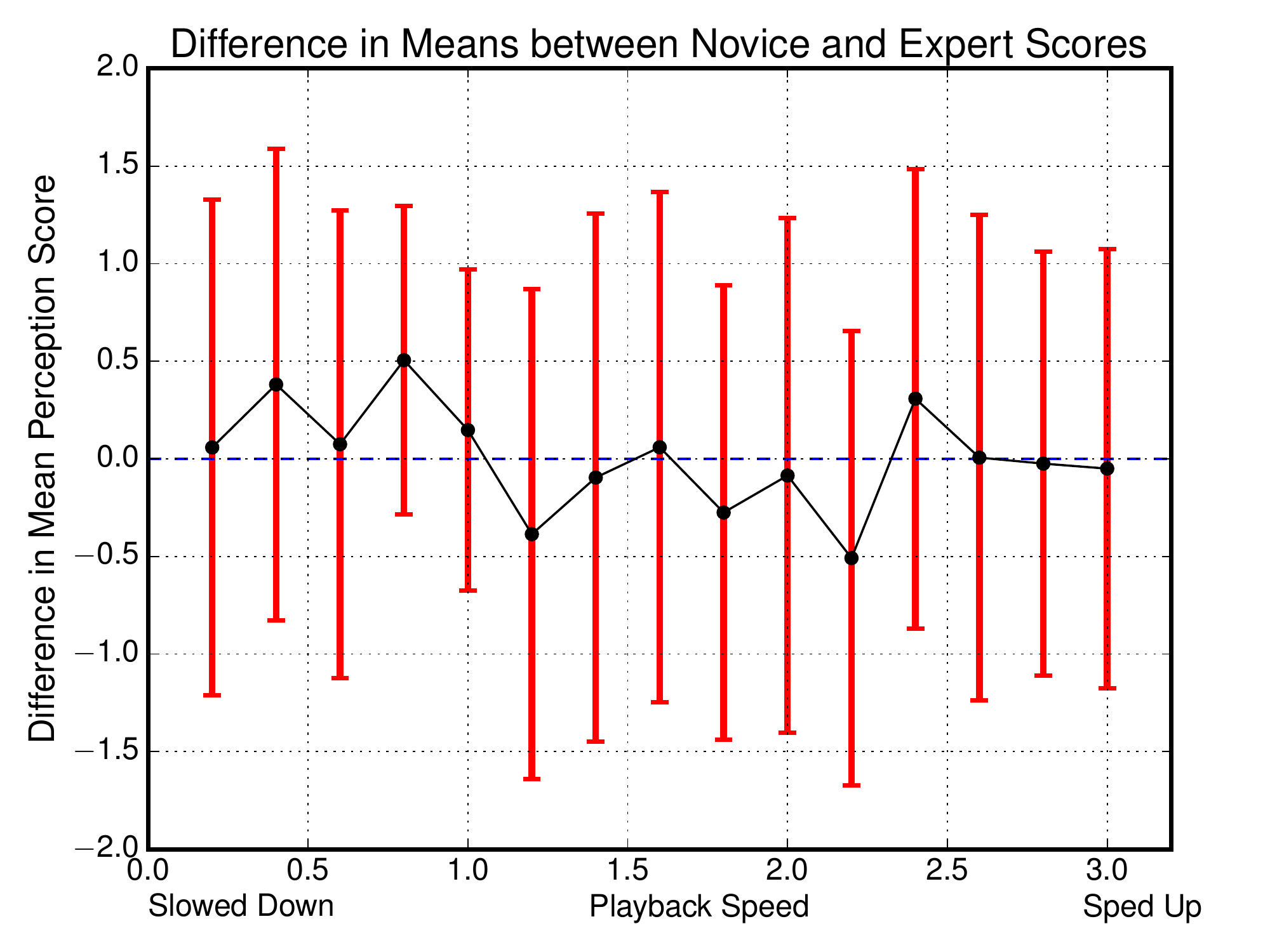}}
        \caption{Difference in Mean between Expert and Novice evaluations.}
    \label{fig:TTest}
   
    \end{subfigure}
    \label{fig:Compare_Speed_GOALS}
     \caption{Comparison of Novice and Expert Speed Perception at each playback speed. (There were $N = 40 \; unique \; human\; evaluations / skill\; level / playback\; speed$. The error bars represent 95\% confidence intervals.)}
\end{figure}

\begin{figure}[h!]
    \begin{subfigure}[t]{\textwidth}
    \centering
        \raisebox{-\height}{\includegraphics[width=0.61\textwidth]{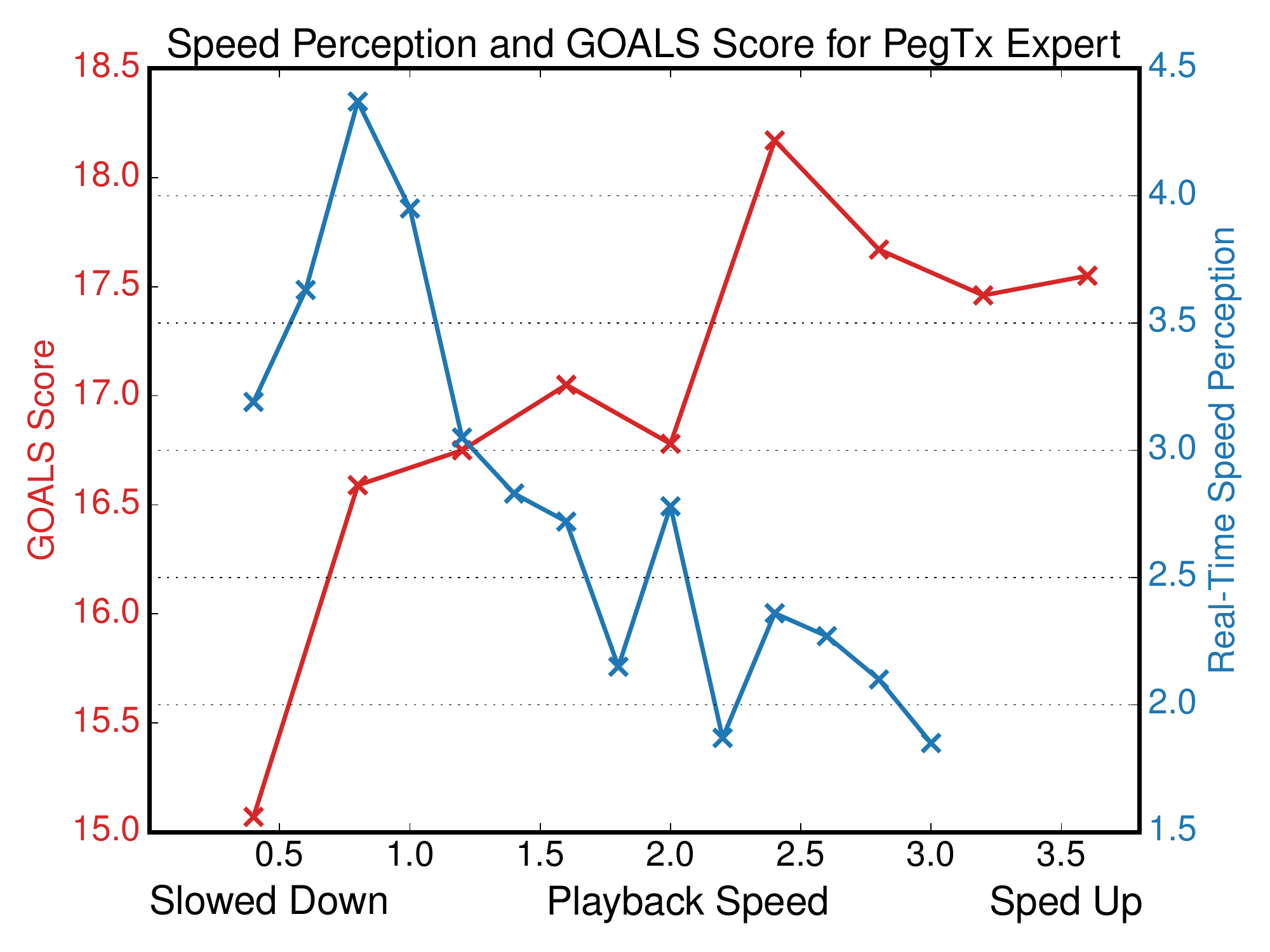}}
        \caption{Expert}
    \end{subfigure}
    \hfill
    \begin{subfigure}[t]{\textwidth}
    \centering
        \raisebox{-\height}{\includegraphics[width=0.62\textwidth]{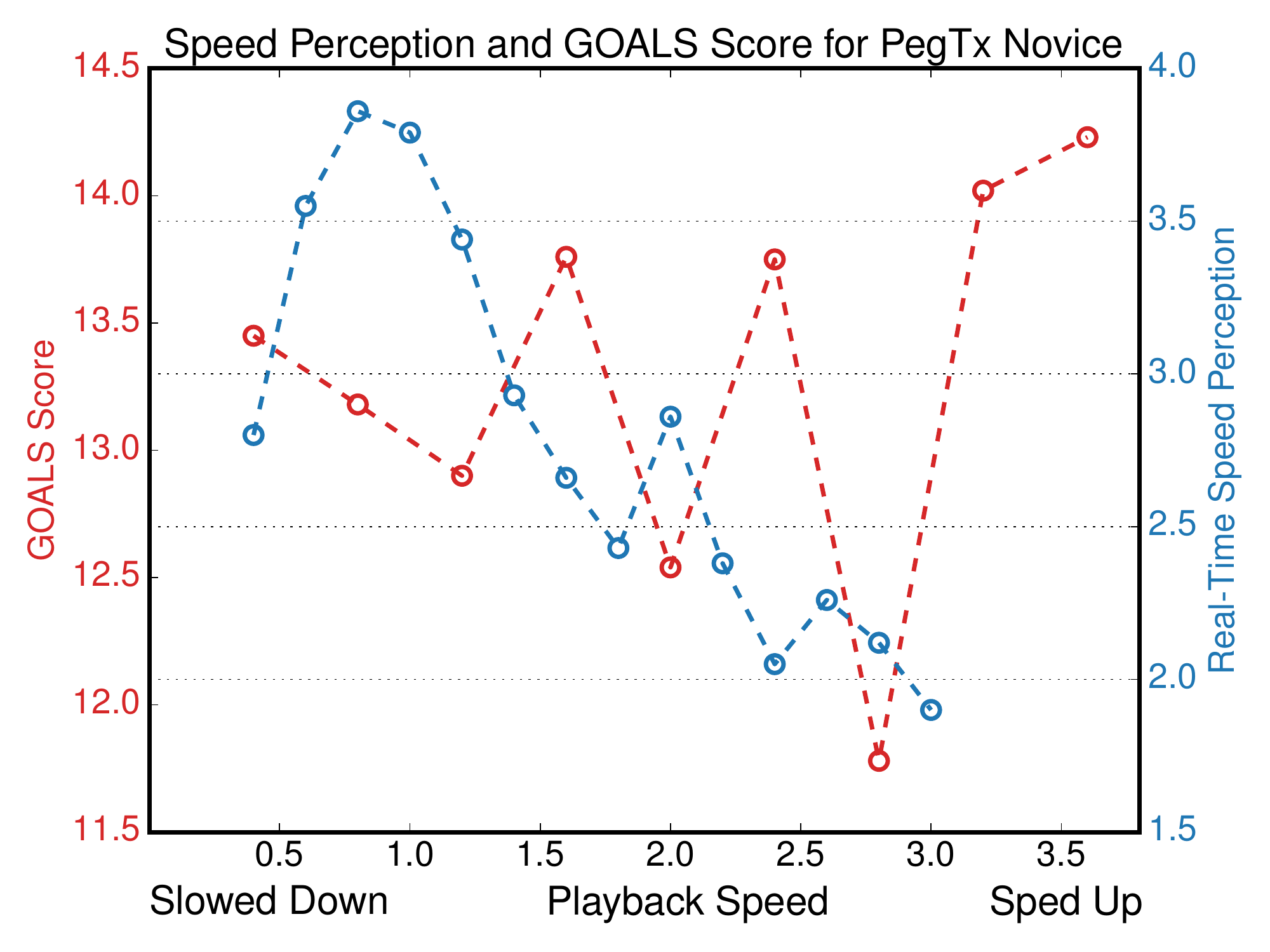}}
        \caption{Novice}
    \end{subfigure}
    \caption{Speed Perception compared with technical skill perceived at various speeds.}
    \label{fig:Compare_Speed_GOALS}
\end{figure}

\subsection{Playback Speed Perception}

The speed perception results for an obvious novice and obvious expert are plotted in Fig. \ref{fig:Speed_Perception_Scores}. The scores were obtained by treating each of the scores in Table \ref{tab:speed_perception_fast} as a score, and obtaining the mean score for each video evaluated, such that a higher score leads to a higher perception of the video being played at a real-time playback speed. Both novices and experts displayed similar perceptions of the video either having an increased or decreased video playback speed. In Fig. \ref{fig:TTest}, a two sample independent t-test was computed between the two groups at each of the playback speeds, to visualize the difference in mean speed perception between experts and novices. All of the speeds show no significant difference in playback speed recognition. The perceived video playback speed scores are compared with the same video's crowd scores for the GOALS assessment metric in Fig. \ref{fig:Compare_Speed_GOALS}.

\section{Conclusion}

The results from the technical skill perception study give support to our initial hypothesis that increasing the video playback speed would increase the ratings of surgical performances. Surprisingly, however, we discovered that novice performances receive a much lower increase in score, which is almost negligible. This finding elucidates the notion that despite increasing the video playback speed of a novice performance, non-expert crowd workers are still able to spot the more obvious mistakes made by these novices. However, it appears that crowd evaluators are biased to speed when they evaluate expert performances. This could be due to expert performers appearing as though their movements are even smoother at quicker playback speeds and the few mistakes they are making being `washed out' or not emphasized at quicker playback speeds. With 135,000 surgeons in the U.S. and the growing need to objectively quantify surgical skills to ensure public safety, methods to rapidly triage technical skill are required. Our observations may provide an easier and cheaper way of discriminating novices from experts than with using expert surgeon evaluation. 

To illustrate whether the `Efficiency' domain of the GOALS assessments was the only domain increasing, as this domain is most closely related to speed, a scatter plot for the average efficiency domain and average of the other three domains for novices and experts is shown in Fig. \ref{fig:Stacked_Area}. The novices clearly show no difference in Efficiency vs. the other three domains. In addition, the three remaining domains for experts appear to increase at almost the same rate as the Efficiency domain. It appears that crowds are also biased to give higher ratings in other domains which aren't necessarily related to speed. 

The results from the real-time playback perception study validate our initial hypothesis by showing us that crowds are able to accurately discern when a video has been manually edited to have a different playback speed. Although, as shown in Fig. \ref{fig:Speed_Perception_Scores}, there is apparently no difference in perception between an obvious novice performance and an obvious expert performance. 

As shown in our comparison to objective metrics, no other used method of objectively obtaining technical skills scores will show the same amount of improvement for increases in playback speed. There may be information that humans can decipher which objective metrics or machine learning algorithms cannot. 

We conclude that increasing the video playback speed of performances of dry lab laparoscopic training tasks could provide a cheap and easy way to discriminate experts from novices as the separation in GOALS scores between novices and experts appears to increase as video playback is increased. A limitation of this study is that we only sampled ten videos of laparoscopic training procedures. Additional investigation with a larger dataset of more clinically relevant performances is required to conclude whether this observation extends to actual surgical case evaluation. 
%

\section*{Compliance With Ethical Standards}
	
	\noindent\textbf{Conflicts of Interest}
	The authors declare that they have no conflict of interest.
	
	\noindent\textbf{Ethical Standard}
	All procedures performed in studies involving
	human participants were in accordance with the ethical standards of
	the institutional and/or national research committee and with the 1964
	Helsinki declaration and its later amendments or comparable ethical
	standards.
	
	\noindent\textbf{Funding}
	This work was supported, in part, by the Office of the Assistant Secretary of Defense for Health Affairs under Award No. W81XWH-15-2-0030, the National Science Foundation CAREER grant under Award No. 1847610, as well as the National Institutes of Health’s National Center for Advancing Translational Sciences, grant UL1TR002494. Opinions, interpretations, conclusions, and recommendations are those of the authors and are not necessarily endorsed by the Department of Defense, the National Science Foundation, or the National Institutes of Healths's National Center for Advancing Translational Sciences.

	\noindent\textbf{Informed Consent} Informed consent was obtained from all individual participants included in the study.

\bibliographystyle{spbasic}      

\begin{thebibliography}{19}
	\providecommand{\natexlab}[1]{#1}
	\providecommand{\url}[1]{{#1}}
	\providecommand{\urlprefix}{URL }
	\expandafter\ifx\csname urlstyle\endcsname\relax
	\providecommand{\doi}[1]{DOI~\discretionary{}{}{}#1}\else
	\providecommand{\doi}{DOI~\discretionary{}{}{}\begingroup
		\urlstyle{rm}\Url}\fi
	\providecommand{\eprint}[2][]{\url{#2}}
	
	
	\bibitem{CSATS_Paper}
	Kowalewski T, Comstock B, Sweet R, Schaffhausen C, Menhadji A, Averch T, Box G, Brand T, Ferrandino M, Kaouk J, Knudsen B, Landman J, Lee B, Schwartz BF, McDougall E, and Lendvay TS (2015)
	Crowd-Sourced Assessment of Technical Skills for Validation of Basic Laparoscopic Urologic Skills (BLUS) Tasks. The Journal of Urology 195(6):1859--1865.	
	
	\bibitem{ThirdLeadingDeath}
	Makary M and Daniel M (2016)
	Medical error - the third leading cause of death in the US. The BMJ.
	
	\bibitem{Birkmeyer}
	Birkmeyer JD, Finks JF, O'Reilly A, Oerline M, Carlin AM, Nunn AR, Dimick J, Banerjee M, and Birkmeyer NJ (2013)
	Surgical skill and complication rates after bariatric surgery. New England Journal of Medicine 369(15):1434--1442.	
	
	\bibitem{MAC}
	Dockter R, Lendvay TS, Sweet RM, and Kowalewski TM (2017)
	The minimally acceptable classification criterion for surgical skill: intent vectors and separability of raw motion data. The International Journal of Computer-Assisted Radiology and Surgery 12:1151--1159.
	
	\bibitem{CSATS}
	Chen C, White L, Kowalewski T, Aggarwal R, Lintott C, Comstock B, Kuksenok K, Aragon C, Holst D, and Lendvay T (2013)
	Crowd-Sourced Assessment of Technical Skills: a Novel Method to Evaluate Surgical Performance. Journal of Surgical Research 187(1):65--71. 
	
	
	\bibitem{GOALS_Paper1}
	Vassiliou MC, Feldman LS, Andrew CG, Bergman S, Leffondre K, Stanbridge D, and Fried GM (2005)
	A global assessment tool for evaluation of intraoperative laparoscopic skills. The American Journal of Surgery 190(1):107--113.
	
	\bibitem{GOALS_Paper2}
	Gumbs AA, Hogle NJ, and Fowler DL (2007)
	Evaluation of Resident Laparoscopic Performance Using Global Operative Assessment of Laparoscopic Skills. Journal of the American College of Surgeons 204(2):308--313.
	
	\bibitem{BeyondTaskTime}
	Kowalewski TM, White LW, Lendvay TS, Jiang IS, Sweet RS, Wright A, Hannaford B, and Sinanan MN (2014)
	Beyond task time: automated measurements augments fundamentals of laparoscopic skills methodology. Journal of Surgical Research 192(2): 329--338. 
	
	\bibitem{Hung}
	Hung A, Chen J, Che Z, Nilanon T, Jarc A, Titus M, Oh PJ, Gill IS, and Liu Y (2018)
	Utilizing Machine Learning and Automated Performance Metrics to Evaluate Robot-Assisted Radical Prostatectomy Performance and Predict Outcomes. Journal of Endourology 32(5). 
	
\bibitem{PLW}
	Barclay CD, Cutting JE, and Kozlowski LT (1978)
	Temporal and spatial factors in gait perception that influence gender recognition. Percpetion and Psychophysics 23(2):145--152.
	
\bibitem{altered_PLW}
	Veto P, Einhauser W, and Troje NF (2017)
	Biological motion distorts size perception. Scientific Reports 7(42576):1--6.
	
\bibitem{speed_bioMotion}
	Jacobs A, Pinto J, and Shiffrar M (2004)
	Experience context and the visual perception of human movement. Journal of Experimental Psychology 30(5):822--835.
	

	\bibitem{Initial_BLUS_curriculum}
	Seete RM, Beach R, Sainfort F, Gupta P, Reihsen T, Poniatowski LH, and McDougall EM (2012)
	Introduction and validation of the American Urological Association Basic Laparoscopic Urology Surgery skills curriculum. Journal of Endourology, 26: 190.
	
	
	\bibitem{Jerk_Hogan} 
	Flash T and Hogan N (1985)
	The Coordination of Arm Movements: An Experimentally Confirmed Mathematical Model. The Journal of Neuroscience 5(7):1688-1703.
	
	\bibitem{Locations_BLUS}
	Kowalewski TM, Seet R, Lendvay TS, Menhadji A, Averch T, Box G, Brand T, Ferrandino M, Kaouk J, Knudsen B, Landman J, Lee B, Schwartz BF, and McDougall E (2016)
	Validation of the AUA BLUS Tasks. Journal of Urology 195: 998.
	
	\bibitem{Omalley_SAL} 
	Bajcsy A, Losey DP, O`Malley MK, and Dragan AD (2018)
	Learning from Physical Human Corrections, One Feature at a Time. In Proceedings of 2018 ACM/IEEE International Conference on Human-Robot Interaction, NY, USA.
	
	\bibitem{Balas_SAL_Creator}
	Balasubramanian S, Melendez-Calderon A, and Burdett E (2012)
	A Robust and Sensitive Metric for Quantifying Movement Smoothness. IEEE Transactions on Biomedical Engineering 59(8):2126--2136.
	
	\bibitem{Kowalewski_Thesis}
	Kowalewski TM (2012)
	Real-time quantitative assessment of surgical skill. PhD Thesis, University of Washington. 
	
	\bibitem{FLS1}
	Derossis AM, Fried GM, Abrahamowicz M, Sigman HH, Barkun JS, and Meakins JL (1998)
	Development of a model for training and evaluation of laparoscopic skills. American Journal of Surgery 175: 482.
	
	\bibitem{FLS2}
	Fried GM (2008) 
	FLS assessment of competency using simulated laparoscopic tasks. Journal of Gastroenterology Surgery 12: 210. 
		
	\bibitem{FLS3}
	Peters JH, Fried GM, Swanstrom LL, Soper NJ, Silin LF, Schirmer B, and Hoffman K (2004)
	Development and validation of a comprehensive program of education and assessment of the basic fundamentals of laparoscopic surgery. Surgery, 135: 21.
	
	\bibitem{FFMPEG}
	FFmpeg Developers (2016)
	ffmpeg tool (Version 4.1.3) [Software]. Available form http://ffmpeg.org/.
	
	\bibitem{Holoborodko}
	Holoborodko P (2008)
	Smooth Noise Robust Differentiators. http://www.holoborodko.com/pavel/numerical-methods/numerical-derivative/smooth-low-noise-differentiators/
	
	\bibitem{Python}
	Python Software Foundation. Python Language Reference, version 3.6. Available at http://www.python.org	
	
	\bibitem{MATLAB}
	MATLAB R2019a, The Mathworks Inc. Natick, Massachusetts.
	
	
	
\end{thebibliography}


\end{document}